\documentclass[aps,prl,twocolumn,superscriptaddress,showpacs,preprintnumbers,amsmath,amssymb]{revtex4}

\usepackage{graphicx}
\graphicspath{{fig/}}
\usepackage{dcolumn}
\usepackage{bm}

\begin{document}

\preprint{APS/123-QED}

\title{Intermediate regimes in granular Brownian motion: Superdiffusion and subdiffusion}

\author{Anna Bodrova}%
\affiliation{%
Faculty of Physics, M.V.Lomonosov Moscow State University, Moscow, 119991, Russia
}%
\author{Awadhesh Kumar Dubey}%
\affiliation{%
School of Physical Sciences, Jawaharlal Nehru University, New Delhi, 110067, India.
}%
\author{Sanjay Puri}%
\affiliation{%
School of Physical Sciences, Jawaharlal Nehru University, New Delhi, 110067, India.
}%
\author{Nikolai Brilliantov}%
\affiliation{%
Department of Mathematics, University of Leicester, Leicester LE1 7RH, United Kingdom
}%

\date{\today}

\begin{abstract}
Brownian motion in a granular gas in a homogeneous cooling state is studied theoretically and by means
of molecular dynamics. We use the simplest first-principle model for the impact-velocity dependent
restitution coefficient, as it follows for the model of  viscoelastic spheres. We reveal that for a wide
range of initial conditions the ratio of granular temperatures of Brownian and bath particles
demonstrates complicated non-monotonous behavior, which results in 
transition between different regimes of Brownian dynamics:~It starts from the ballistic motion, switches
later to superballistic one and turns at still later times into subdiffusion; eventually normal
diffusion is achieved. Our theory agrees very well with the MD results, although extreme computational
costs prevented to detect the final diffusion regime. Qualitatively, the  reported intermediate
diffusion regimes  are generic for granular gases with any realistic dependence of the restitution
coefficient on the impact velocity.
\end{abstract}


\maketitle

\textit{Introduction.} Brownian motion is a fundamental process in nature which takes place on very
different time and length scales~\cite{Brownbook}. In its classical formulation it implies a random
motion of a big (Brownian) particle driven by thermal motion of much smaller bath particles. The
mean-square displacement of Brownian particles grows with time as $\left\langle R_B^2\right\rangle \sim
t^{\beta}$, with $\beta =1$ for normal diffusion. Anomalous diffusion with  $\beta>1$ (superdiffusion)
and $\beta <1$ (subdiffusion) has been found  in a variety of systems~\cite{Andif}, ranging from
supercooled and glass forming liquids, e.g.~\cite{glas}, to surfaces~\cite{expot},
biological~\cite{crowd} and granular~\cite{angran} systems; usually a few diffusion regimes with a
crossover between them are observed. Mechanisms  of anomalous diffusion are seemingly very different --
macromolecular crowding in solutions and cells~\cite{crowd}, caging in glassy liquids~\cite{glas},
trapping in a random potential~\cite{expot}, etc. Nevertheless, one can possibly infer an important
common feature -- anomalous diffusion always arises due to interplay of different basic dynamics in a
system:~For instance, a (fast) hopping dynamics in glass forming  liquids interferes with a (slow)
dynamics of structural rearrangements, resulting in subdiffusion. Moreover, anomalous diffusion in most
of the systems manifests as an intermediate regime, which transforms asymptotically into normal
diffusion, e.g.~\cite{Andif,glas,expot}.

In the present Letter we report intermediate anomalous diffusion in granular systems, where particles
interact with dissipative forces.  Although the microscopic mechanism of this phenomenon is novel  and
related to inelastic collisions,  it again results,  as it is shown below,  from the interplay of
different dynamics represented by evolution of two granular temperatures.

Granular  Brownian motion demonstrates a large variety of  new surprising phenomena,
among which is the violation of energy
equipartition in a mixture of massive and light
particles~\cite{book,breydufty,GarzoDuftyMixtureHrenyaWildmanMenon}: Both granular temperatures of
Brownian  ($T_B$) and bath ($T$) particles decrease in a force-free granular gas, while their  ratio
rapidly relaxes to a steady-state value $(T_B/T)_{\rm s.s.}>1$. Moreover, $(T_B/T)_{\rm s.s.}$ increases
with increasing inelasticity and the  mass ratio $m_B/m$ of Brownian ($m_B$) and bath ($m$)
particles~\cite{book,breydufty}. It has been also shown that Brownian motion in these systems may change
qualitatively  --  from  diffusive to a ballistic one, depending on the steady state value of  $\varphi
= \left(T_Bm\right)/\left(Tm_B\right)$~\cite{pht}. Surprisingly, one can treat this phenomena as a
"phase transition" with the "order parameter" $\varphi$~\cite{pht}, which however is not observed as a
real process, since  the steady state $\varphi$ is constant.

To date, the most theoretical studies of Brownian motion are done with the assumption of a constant
restitution coefficient $\varepsilon = g\prime/g$~\cite{book}, where $g$ and $g\prime$ are,
respectively, the normal components of the relative inter-particle velocities before and after a
collision. The simplifying hypothesis of constant $\varepsilon$, however, contradicts experimental
results, which indicate that $\varepsilon$ does depend on $g$ and that $\varepsilon(g)$ is a decreasing
function of the impact velocity~\cite{goldsmitBridgesHatzesLin:1984}; it is not also consistent with a
rigorous theoretical analysis~\cite{book}. The impact-velocity dependence of the restitution coefficient
may, however, crucially change the behavior of a granular system, e.g.~\cite{vdvephysa}. In particular,
it has been shown that in a gas of viscoelastic particles, where the impact-velocity dependence of
$\varepsilon$ stems from the simplest first-principle model~\cite{bril96ramirez}, clusters and vortices
manifest as transient phenomena~\cite{BrilPRL2004}, in a sharp contrast with the case of a constant
$\varepsilon$~\cite{clusters}. Furthermore, in a mixture of massive and light particles, the temperature
ratio $T_B/T$, as well as $\varphi$, evolve with time in a HCS~\cite{jetp}, again in a contrast with the
case of $\varepsilon = \rm const$. Therefore, one can expect that in a granular gas of viscoelastic
particles, where  $\varphi$ varies with time, the "phase transition" between different regimes of
Brownian motion may occur during the gas evolution as a real physical process.

We analyze granular Brownian  motion in a gas of viscoelastic particles in a homogeneous cooling state
(HCS) -- the basic state of a force free granular gas by means of molecular dynamics (MD) and
theoretically. We have revealed a sequence of new diffusive regimes for a Brownian particle --
superballistic motion (superdiffusion), which follows after ballistic one and transforms then to
subdiffusion. Eventually, normal diffusion is observed.

\textit{MD simulations.} We perform event-driven simulations~\cite{CompBook}  for a force-free system,
using  $N=64000$ bath particles of mass $m$ and diameter $\sigma$ and one Brownian particle of mass $m_B
\gg m$ and diameter $\sigma_B$ (for simplicity we take $\sigma_B = \sigma$) in a cube of length $L =
130\,\sigma$ with periodic boundary conditions. The number density of the bath particles  was $n =
N\pi\sigma^3/(6L^3) \simeq 0.015$. The reported results correspond to averages over 20 or 50 independent
runs, depending on the initial conditions.

\begin{figure}
\includegraphics[width = 0.83\columnwidth]{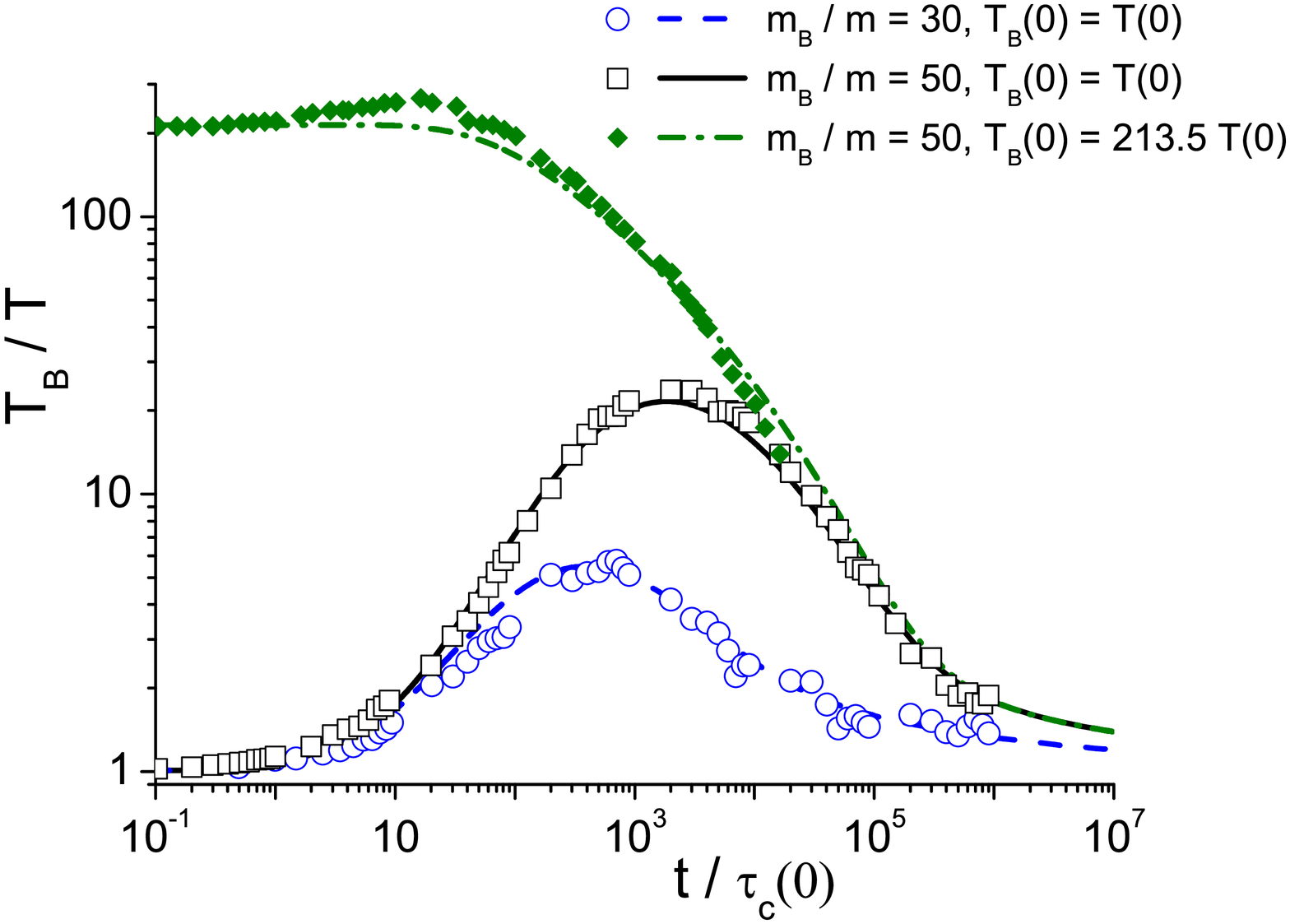}
\includegraphics[width = 0.83\columnwidth]{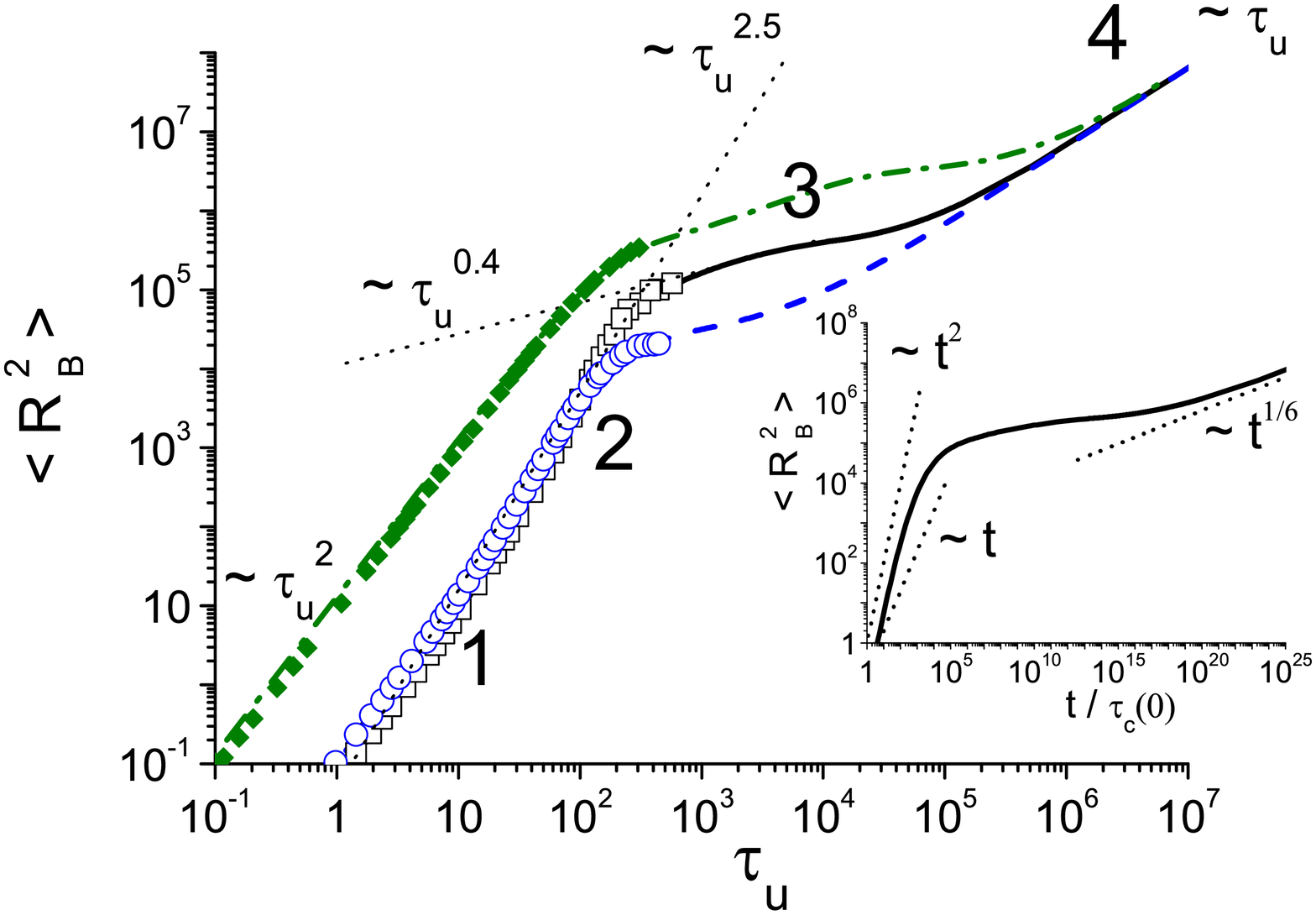}
\caption{\textbf{\textit{(a)}} Time dependence of the  temperature ratio of Brownian ($T_B$) and bath
($T$) particles. \textbf{\textit{(b)}} The mean-square displacement of Brownian particles $\left\langle
R_B^2\right\rangle$ as the function of the re-scaled time $\tau_u$ and the laboratory time $t$ (inset).
Lines -- theory, symbols -- MD data. Numbers indicate different regimes:~1 - ballistic motion,
$\left\langle R^2_B\right\rangle \sim \tau_u^2$;  2 - superballistic regime (superdiffusion),
$\left\langle R^2_B\right\rangle \sim \tau_u^{\beta}$ with $\beta
> 1$; 3 - subdiffusion, $\left\langle R^2_B\right\rangle \sim \tau_u^{\beta}$ with  $\beta <1 $; 4 -
normal diffusion $\left\langle R^2_B\right\rangle \sim \tau_u $. }
 \label{GTbT}
\end{figure}

In our simulations we use the restitution coefficient, as it follows for the model of viscoelastic
particles~\cite{bril96ramirez,delayed}:
\begin{equation}
\varepsilon=1+\sum_{k=1}^{\infty}C_{k}\delta^{k/2}\left(2u\right)^{k/20}w^{k/10} \, .
\label{eps1}
\end{equation}
Here, $C_{k}$'s are numerical coefficients, which have been computed up to $k=20$~\cite{delayed} and
$\delta$ quantifies the dissipative interactions. Eq.~(\ref{eps1}) describes both types of collisions --
between Brownian particle and a bath particle and between the bath particles themselves. In the former
case $w = \left| \left(\vec{c}_{Bi}\cdot\vec{e}\, \right)\right|$, where
$\vec{c}_{Bi}=\vec{v}_{Bi}/v_T$, with $\vec{v}_{Bi} = \vec{v}_B - \vec{v}_i$ being the relative velocity
of Brownian ($\vec{v}_B$) and $i$-th bath ($\vec{v}_i$) particle, $v_T = \sqrt{2T/m}$ is the thermal
velocity of the bath particles and $\vec{e}$ is the unit vector, joining particles' centers at the
collision instant. In the latter case $w = \left| \left(\vec{c}_{ij}\cdot\vec{e}\right)\right|$, where
$\vec{c}_{ij}=\vec{v}_{ij}/v_T$, with $\vec{v}_{ij} = \vec{v}_i - \vec{v}_j$ being the relative velocity
of the colliding $i$-th and $j$-th bath particles. For both types of collisions we use for simplicity
the same $\delta=0.1$.  The quantity $u(t)=T(t)/T(0)$ is the dimensionless temperature of the bath at
time $t$, and  granular temperatures are defined as usual, $3T/2=\left\langle m\vec{v}^{\,
2}/2\right\rangle$ and $3T_B/2=\left\langle m_B\vec{v}_B^{\, 2}/2\right\rangle$ . The results of the MD
simulation are presented in Fig.~\ref{GTbT}.

\textit{Theory.} Since concentration of Brownian particles is much smaller than that of the bath
particles, we assume,  that Brownian particles do not affect  evolution of the bath~\cite {book}: In a
force-free gas $T$ gradually decreases, due to dissipative collisions, with the cooling coefficient
$\xi(t) = -\left(dT/dt\right)/T\left(t\right)$. As temperature $T$ tends to zero, $\varepsilon$ tends to
unity, 
and all collisions tend to be elastic~\cite{book}.

Evolution of Brownian particles may be described, using the Boltzmann equation, $\dot{f}_B(\vec{v}_B,t)
=  I (f_B,f)$ for the velocity distribution function, $f_B(\vec{v}_B,t)$, e.g.~\cite{book}; the
collision integral $I(f_B,f)$  accounts for the alteration of $f_B(\vec{v}_B,t)$ in pair-wise collisions
with the collision rule,
\begin{equation}
\vec{v}_{B}^{\, \prime} = \vec{v}_{B}-\left(1+\varepsilon \right) \mu \left(\vec{v}_{Bi}\cdot\vec{e}\,
\right)\vec{e} \, . \label{velplus}
\end{equation}
Here $\vec{v}_B^{\, \prime}$ is the velocity of a Brownian particle after a collision and $\mu \equiv
\frac{m}{m_B+m}$. Solving the Boltzmann equation with the standard technique, e.g.~\cite{book}, one
obtains the distribution function $f_B$. Brownian dynamics may be described with very compact notations
if the pseudo-Liouville operator $\hat{L}$ is used~\cite{book}:
\begin{equation}
 \dot{\vec{v}}_B= \hat{L} \vec{v}_B = \hat{L}_0\vec{v}_B + \sum_{i}\,
\hat{T}_{Bi}\vec{v}_B\,,  \label{eq:eq2}
\end{equation}
where $\hat{L}_0=\vec{v}_B \! \cdot \!  \vec{\nabla}_{\vec{r}_B}$ describes the free streaming and
\begin{equation}
\hat{T}_{Bi} \!=\! \sigma_{0}^2 \!\int \!d\vec{e}
\Theta\left(-\left(\vec{v}_{Bi}\cdot\vec{e}\right)\right)\left|\vec{v}_{Bi}\cdot\vec{e}\right|
\delta\left(\vec{r}_{Bi}\!-\!\vec{e}\sigma_0\right)(\hat{b}_{Bi}-1) \nonumber
\end{equation}
the binary collisions of particles. The radius-vector $\vec{r}_{Bi}$ in the above equation joins centers
of Brownian and bath particle,  $\sigma_{0} = \left(\sigma_B+\sigma\right)/2$ and the operator
$\hat{b}_{Bi}$ acts on the velocities, as $\hat{b}_{Bi}\vec{v}_{B} = \vec{v}_{B}^{\, \prime}$ with the
collision rule (\ref{velplus}).

Taking into account, that $\vec{R}_B(t) = \int_0^{t} \vec{v}_B(t_1) d t_1$, the mean-square displacement
of Brownian particles reads in terms of the velocity autocorrelation function: $\left\langle
R_B^2(t)\right\rangle = \int_0^t dt_1\int_0^t dt_2 \left\langle
\vec{v}_B\left(t_1\right)\vec{v}_B\left(t_2\right) \right\rangle$. We introduce the reduced time
$\tau_B$, defined as $d\tau_B = dt \sqrt{T_B(t)/T_B(0)}/\tau_c(0)$, where
$\tau_{c}^{-1}(t)=4\sqrt{\pi}\sigma^{2}g_{2}(\sigma)n\sqrt{T(t)/m}$ is the mean collision time of bath
particles, with $g_2(\sigma)$ being the contact value of the pair correlation function~\cite{book}. In
this time scale one deals with the reduced velocities, $\vec{c}_B = \vec{v}_B/\sqrt{2T_B/m_B}$, and
granular temperature of Brownian particles remains constant. Hence the decay of the (reduced) velocity
correlation function occurs similarly, as in equilibrium gases: A particle loses memory of its initial
velocity in random collisions; in dilute gases the velocity correlation function is
exponential~\cite{Resibois}, therefore we approximate
\begin{equation}
\left\langle \vec{c}_B\left(\tau_B(t_1)\right)\vec{c}_B\left(\tau_B(t_2)\right)\right\rangle \!=
\!\left< c_B^2 \right>
\exp\!\left[-\frac{\tau_B(t_2)-\tau_{B}(t_1)}{\hat{\tau}_{vB}\left(t_1\right)}\right], \label{fcor}
\end{equation}
where $\left< c_B^2 \right> = 3/2$ and $\hat{\tau}_{vB}$ is the (reduced) velocity relaxation time.
Exploiting the above form of the correlation function, we obtain the mean-square displacement:
\begin{equation}
\left\langle R^2_B(t)\right\rangle \!= \!6 \! \int_0^{t}\! dt_1
D_B(t_1)\!\left[1-\exp\left(-\frac{\tau_B(t)-\tau_{B}(t_1)}{\hat{\tau}_{vB}\left(t_1\right)}\right)\right].
\nonumber
 \label{eR2b}
\end{equation}
Here $D_B= T_B(t)\tau_{vB}(t)/m_B$ is the time-dependent diffusion coefficient of a Brownian particle,
expressed in terms of $\tau_{vB}(t) = \hat{\tau}_{vB}(t)\tau_c(0)/\sqrt{T_B(t)/T_B(0)}$ -- the velocity
relaxation time in  laboratory time units.  The reduced relaxation time $\hat{\tau}_{vB}(t) $ in the
exponential function in Eq.~(\ref{fcor}) may be easily found from its time derivative at $t_1 \to t_2$,
yielding~\cite{book,SDbrey1}
\begin{equation}
\tau_{vB}^{-1} =
-(N-1)\frac{\langle\vec{v}_{b}\hat{T}_{Bi}\vec{v}_{b}\rangle}{\langle\vec{v}_{b}^{\,2}\rangle} -
\frac{1}{2}\xi_B \, , \label{tauvb}
\end{equation}
with  the cooling coefficient  of Brownian particles
\begin{equation}
\label{xib} \xi_B = - \frac{1}{T_B}\frac{dT_B}{dt} = -(N-1)\frac{\left\langle
\hat{T}_{Bi}v_B^2\right\rangle}{\left\langle v_B^2\right\rangle} \, .
\end{equation}
Performing the averaging in  Eqs.~(\ref{tauvb}) and (\ref{xib}), we ignore, for simplicity, deviations
of the velocity distribution function from the Maxwellian~\cite{vdvephysa}; it may be however shown that
their impact on the calculated quantities is negligible. Hence, we obtain the velocity relaxation time
\begin{equation}
\nonumber \frac{\tau_{vB}^{-1}(t)}{\tau_c^{-1}(0)} =  \sqrt{\frac{8u}{9}}
\frac{\mu^2(1+\varphi)^{3/2}}{\varphi} \frac{\sigma_0^2g_2\left(\sigma_0\right)}{\sigma^2g_2
\left(\sigma\right)} \left[1+\frac{1}{2}\sum_{i=2}^{\infty}A_iB_i \right]
\end{equation}
and the cooling coefficient of Brownian particles:
\begin{eqnarray}
\nonumber \xi_B(t) &=& 2\tau_c^{-1}(0) \sqrt{\frac{8u}{9}} \left(1+\varphi\right)^{1/2}\mu
\frac{\sigma_0^2g_2\left(\sigma_0\right)}{\sigma^2g_2\left(\sigma\right)}\times
\\ \nonumber
&\times&\left[1-
 \mu \frac{1+\varphi}{\varphi}+\sum_{i=2}^{\infty}B_i\left(C_i- \frac12 \mu
 \frac{1+\varphi}{\varphi}A_i\right)
\right] \, ,\label{xibfull}
\end{eqnarray}
where $A_i=4C_i+\sum_{j+k=i}C_jC_k$ are pure numbers and
\begin{equation}
B_i(t)=\delta^{i/2}\left(2u(t)\right)^{i/20}(1+\varphi(t))^{i/20} \left(\frac{20i+i^2}{800}\right) \Gamma\left(\frac{i}{20}\right)
\nonumber
\end{equation}
with $\Gamma\left(x\right)$ being the Gamma-function.

The temperature of the  Brownian particles $T_B$ can be found from  the equation $dT_B(t)/dt = -\xi_B(t)
T_B(t)$, while the temperature $T(t)$ and the mean-squared displacement $\left\langle R^2(t)
\right\rangle$ of the bath particles -- from  the above results using  the substitute $m_B \rightarrow
m$ and $\sigma_B \rightarrow \sigma$.

\textit{Results and discussion.} The theoretical predictions are compared with the simulation data in
Fig.~\ref{GTbT}. First we consider the case of the initial energy equipartition, $T_B(0) = T(0) =
400/3$. The time dependence of the ratio $T_B/T$  demonstrates here a complicated non-monotonous
behavior: It initially increases, reaches a maximum, and then decreases, tending eventually to unity,
Fig.~\ref{GTbT}a. This is in a sharp contrast with the  case of a constant $\varepsilon$, where $T_B/T$
rapidly relaxes to a steady-state value. One can explain this effect as follows. From the collision
rules [see, e.g. Eq.~(\ref{velplus})],  the ratio of energy losses of Brownian ($\Delta E_B$) and bath
($\Delta E$) particles at a collision, scales as $\Delta E_B/\Delta E \sim v_B/v_i \sim
\sqrt{(T_Bm)/(Tm_B)}$. Initially $\Delta E_B  \ll \Delta E$, since $T_B=T$ and $m_B \gg m$, therefore
Brownian particles cool down significantly slower than the bath particles and the temperature ratio
$T_B/T$ increases. One can call this effect as a "retarded cooling", which may be also understood as a
tendency of a system to reach the  ratio of $T_B/T>1$, corresponding to the steady-state value
$(T_B/T)_{\rm s.s.}>1$ of a gas with the respective constant $\varepsilon$. The larger the mass ratio
the more pronounced the effect (Fig.~\ref{GTbT}a). In the course of time the temperature ratio $T_B/T$
gets so large that $\Delta E_B$ and $\Delta E$ become comparable; the granular temperatures start then
to equilibrate. At this stage $T_B/T$ decreases -- Brownian particles cool faster than the bath
particles and one can call this regime as "accelerated cooling". Again, this effect may be understood,
noticing that the ratio of $(T_B/T)_{\rm s.s.}>1$ is smaller for larger $\varepsilon$ and that the
effective restitution coefficient grows as the gas cools down, see Eq.~(\ref{eps1}). For very large
times $\varepsilon$ tends to unity, the system becomes elastic and the energy equipartition is achieved.
 This complicated temperature dependence results in transition between different diffusion regimes,
which may be most clearly seen using the re-scaled time $\tau_u$, defined as $d\tau_u =
dt\sqrt{u(t)}/\tau_c(0)$. In this time scale the average velocity of the bath particles is  constant and
they move, as in an equilibrium molecular gas, that is, ballistically,  $\left\langle R^2\right\rangle
\sim \tau_u^2$,  for $\tau_u \sim 1$ and diffusively, $\left\langle R^2\right\rangle \sim \tau_u$,  for
$\tau_u  \gg  1$ \cite{book,SDbrey1}. Brownian motion demonstrates, however, four different regimes --
ballistic (regime 1 in Fig.~\ref{GTbT}b), superballistic motion or superdiffusion (regime 2),
subdiffusion (regime 3) and normal diffusion (regime 4). Physically, the superballistic motion, which
starts after a short ballistic regime corresponds to the "retarded cooling", while the subsequent
subdiffusion -- to the "accelerated cooling". In the former case Brownian particles get hotter and
hotter with respect to the surrounding gas and  $\left\langle R^2_B\right\rangle$ grows with time faster
than for normal diffusion; in the latter case, Brownian particles cool more rapidly than the gas, so
that $\left\langle R^2_B\right\rangle$ grows with time slower. The duration of the subdiffusion regime
increases with the mass ratio $m_B/m$  and may be very long. Asymptotically, at $\tau_u \to \infty$, the
system returns to the equipartition, demonstrating the normal diffusion, $\left\langle
R^2_B\right\rangle \sim \tau_u$.

The respective transition between diffusion regimes is also seen for the laboratory time $t$
(Fig.~\ref{GTbT}b, inset), where the "normal diffusion" is described by  $\left\langle
R^2_B\right\rangle \sim t^{1/6}$, as it follows for the gas of viscoelastic
particles~\cite{book,SDbrey1}. If we consider the time scale $\tau_B$, where $T_B$ keeps constant, the
superballistic motion does not appear; all three other  diffusion regimes, nevertheless, present.

Since superballistic motion is caused by the "retarded cooling",  it is not observed if $T_B/T$ does not
increase. This happens for initial conditions, when  the ratio $T_B/T$ is already large, or more
precisely, when $T_B/T$ exceeds some  threshold, estimated as  the ratio $(T_B/T)_{\rm s.s.}$ for a gas
with a constant $\varepsilon$,  equal to  the effective restitution coefficient $\varepsilon_{\rm eff}$
for a gas of viscoelastic particles at $t=0$. In our case Eq.~(\ref{eps1}) with $u(0)=1$ and $\left<w
\right>=4/\sqrt{2 \pi}$ yields $\varepsilon_{\rm eff}(0) = 0.876$ and $T_B(0)/T(0)=(T_B/T)_{\rm
s.s.}=213.5$. Hence, this initial temperature ratio  delimits two incipient diffusion regimes:~For
$T_B(0)/T(0) < 213.5$ the ballistic and then superballistic motion takes place, while otherwise only
ballistic regime is observed, Fig.~\ref{GTbT}b. Interestingly, the ballistic motion in the latter case
lasts for significantly  longer time, than for normal diffusion; this follows from the fact that
trajectories of heavy Brownian particles are  almost not affected by much colder bath of light
particles. The subsequent subdiffusion and  normal diffusion persist independently on initial
conditions, see Fig.~\ref{GTbT}b. This is the universal feature of Brownian motion in a gas of
viscoelastic particles. As it follows from Fig.~\ref{GTbT}, a very good agreement between our theory and
simulation data is observed for  evolution of temperature and  mean-square displacement for intermediate
ballistic, superballistic and the onset of subdiffusive regimes; due to extreme computational costs we
fail to detect numerically the final regime of normal diffusion.

 \textit{Conclusion.} We have studied Brownian motion in a force-free granular gas,
 composed of particles interacting with an impact-velocity dependent restitution
 coefficient $\varepsilon(g)$, as it follows from the model of viscoelastic spheres. We have revealed that
 for a wide range of initial conditions the ratio of granular temperatures of massive and light particles
 $T_B/T$ demonstrates complicated non-monotonous dependence on time as the system evolves in the homogeneous cooling
 state. Somewhat similar to other systems with anomalous diffusion, this interplay of different dynamics of Brownian and bath particles, quantified by their
 temperatures, gives rise to a sequence of intermediate diffusive regimes in  granular Brownian motion: At early
 times  the ballistic  motion is observed, which alters then to superballistic one (for non-monotonous
 evolution of $T_B/T$).  At  still later times the opposite regime of subdiffusion -- motion, slower
 than normal diffusion, always takes place; it lasts for a relatively long time and tends asymptotically
 to normal diffusion. Qualitatively, these features of granular Brownian motion are not specific to a
 particular model of $\varepsilon(g)$ but generic for granular systems with any realistic restitution
 coefficient that increases with decreasing impact velocity.

\end{document}